\documentclass[5p,,preprint, 12pt]{elsarticle}

\makeatletter
\usepackage{setspace}
\usepackage{hyperref}
\usepackage{graphicx}
\usepackage{footnote}
\usepackage{tabulary,xcolor}
\usepackage{float}
\usepackage{amsfonts,amsmath,amssymb}
\usepackage{bm}
\usepackage{multirow}
\usepackage{longtable}
\usepackage{breqn}

\journal{Physical Letters B}

\begin{document}
\sloppy
\begin{frontmatter}

\title{Summit of the N=40 Island of Inversion: precision mass measurements and ab initio calculations of neutron-rich chromium isotopes}

\author{R. Silwal\textsuperscript{a,b}, C. Andreoiu\textsuperscript{c}, B. Ashrafkhani\textsuperscript{d}, J. Bergmann\textsuperscript{e}, T. Brunner\textsuperscript{f}, J. Cardona\textsuperscript{a,g}, K. Dietrich\textsuperscript{a, h}, E. Dunling \textsuperscript{a,i}, G. Gwinner\textsuperscript{g}, Z. Hockenbery\textsuperscript{a,f}, J. D. Holt\textsuperscript{a,f}, C. Izzo\textsuperscript{a}, A. Jacobs\textsuperscript{a,j}, A.Javaji\textsuperscript{a,j}, B. Kootte\textsuperscript{a,g}, Y. Lan\textsuperscript{a,j}, D. Lunney\textsuperscript{k}, E. M. Lykiardopoulou \textsuperscript{a,j}, T. Miyagi\textsuperscript{a,l,m}, M. Mougeot\textsuperscript{n,o}, I. Mukul\textsuperscript{a}, T. Murb{\"o}ck\textsuperscript{a}, W. S. Porter\textsuperscript{a,j}, M. Reiter\textsuperscript{p}, J. Ringuette\textsuperscript{a,q}, J.Dilling\textsuperscript{a,j}, and A.A. Kwiatkowski\textsuperscript{a,r}}

\address{\textsuperscript{a} TRIUMF, Vancouver BC V6T2A3 Canada}
\address{\textsuperscript{b}Department of Physics and Astronomy, Appalachian State University, Boone NC 28608 USA}
\address{\textsuperscript{c}Department of Chemistry, Simon Fraser University, Burnaby BC V5A1S6 Canada}
\address{\textsuperscript{d}Department of Physics and Astronomy, University of Calgary, Calgary AB T2N 1N4 Canada}
\address{\textsuperscript{e}Physikalisches Institut, Justus-Liebig-Universite{\"a}t, Gie{\ss}en 35392 Germany}
\address{\textsuperscript{f}Department of Physics, Mcgill University, Montreal QC H3A0G4 Canada}
\address{\textsuperscript{g}Department of Physics and Astronomy, University of Manitoba, Winnipeg MB R3T 2N2 Canada}
\address{\textsuperscript{h}University of Heidelberg, Philosophenweg 12, Heidelberg 69120 Germany}
\address{\textsuperscript{i}Department of Physics, University of York, York YO10 5DD UK}
\address{\textsuperscript{j}Department of Physics and Astronomy, University of British Columbia, Vancouver BC V6T 1Z4 Canada}
\address{\textsuperscript{k}Universit\'e Paris-Saclay, IJCLab-IN2P3/CNRS France}
\address{\textsuperscript{l}Technische Universit\"at Darmstadt, Department of Physics, Darmstadt 64289 Germany}
\address{\textsuperscript{m}ExtreMe Matter Institute EMMI, GSI Helmholtzzentrum f\"ur Schwerionenforschung GmbH, Darmstadt 64291 Germany}
\address{\textsuperscript{n}CERN, 1211 Geneva 23 Switzerland}
\address{\textsuperscript{o} Max-Planck-Institut f\"ur Kernphysik, Heidelberg 69117 Germany} \address{\textsuperscript{p}School of Physics and Astronomy, University of Edinburgh, Edinburgh EH8 9YL UK}
\address{\textsuperscript{q}Department of Physics, Colorado School of Mines, Golden CO 80401 USA}
\address{\textsuperscript{r}Department of Physics and Astronomy, University of Victoria, Victoria BC V8P 5C2 Canada}




\begin{abstract}
Mass measurements continue to provide invaluable information for elucidating nuclear structure and scenarios of astrophysical interest. The transition region between the $Z = 20$ and $28$ proton shell closures is particularly interesting due to the onset and evolution of nuclear deformation as nuclei become more neutron rich. This provides a critical testing ground for emerging ab-initio nuclear structure models. Here, we present high-precision mass measurements of neutron-rich chromium isotopes using the sensitive electrostatic Multiple-Reflection Time-Of-Flight Mass Spectrometer (MR-TOF-MS) at TRIUMF's Ion Trap for Atomic and Nuclear Science (TITAN) facility. Our high-precision mass measurements of $^{59, 61-63}$Cr confirm previous results, and the improved precision in measurements of $^{64-65}$Cr refine the mass surface beyond N=40. With the ab initio in-medium similarity renormalization group, we examine the trends in collectivity in chromium isotopes and give a complete picture of the  N=40 island of inversion from calcium to nickel. 

\end{abstract}

\begin{keyword}
\texttt{mass measurement, MR-TOF-MS, nuclear structure, two neutron separation energies, island of inversion, intruder configuration}
\end{keyword}
\end{frontmatter}


\section{Introduction}
Mass measurements are essential to derive the binding energy of the nucleus needed to understand the nuclear forces that hold the nucleons together \cite{Dilling2018}, and constrain nuclear reaction rates. Such measurements for radioactive nuclei play an important role to enhance our knowledge of nuclear structure, for instance, via the study of evolution of magicity in the nuclear chart. New contenders of nuclei with magic behavior (typically associated with shell gaps in the mass surface, large 2$+$ excitation energies, reduced transition probabilities, among others) were discovered from measurements of nuclei far from stability \cite{Warner2004,  Sorlin2008}. Magic behavior has also been reported for $^{52, 54}$Ca ($Z =20$, $N=32$, $34$) \cite{Steppenbeck2013, Wienholtz2013} and $^{68}$Ni (Z=28, N=40) \cite{Bernas1982} with relatively weak sub-shell closures. Complementing new areas of magicity are regions where expected magic behavior breaks down \cite{Warner2004,  Sorlin2008}. One example is the N=20 islands of inversion (IOI), originally discovered through mass measurements \cite{Sorlin2008, Thibault75}, followed by the N=28 IOI \cite{Sarazin2000, Sorlin2013}. Beyond N=28, the N=40 region has gathered much interest in the last decade as a possible IOI site with signs of strong collective behavior and shape deformation \cite{Gade2021, Gade2010, Michimasa2020, Lungjvall2010, Baugher2012, Mougeot2018, Crawford2013, Braunroth2015}. Some measurements suggest the collective behavior to persist even past N=40 isotones \cite{Santamaria2015, NOWACKI2021}, and with large-scale shell-model calculations predicting even an extension to a fifth IOI \cite{Nowacki2016, Kwiatkowski2015}. Since such an IOI is currently beyond reach of experiment, this prediction should be checked by other theoretical approaches. Therefore, further refining the current ab-initio theories is essential to better describe the changing shell structure. 

For exotic nuclei with a large asymmetry in their proton-neutron ratios, the existing mean-field picture deviate drastically when extrapolated to unmeasured isotopes. These are sometimes attributed to the tensor forces arising from the nucleon-nucleon interactions that shift the single-particle energies in the proton/neutron orbitals \cite{Otsuka2005} creating an IOI. The neutron-rich Cr (Z=24, N=40) isotopic chain that lies midway between the proton shell closures Ca (Z=20) and Ni (Z=28) region is particularly interesting given its extreme proton-neutron ratios. When protons are added to the 1{\it f}$_{7/2}$ orbital, the relative energies of the neutron 2{\it p}$_{1/2, 3/2}$ and 1{\it f}$_{5/2}$ orbitals change due to the strong attractive spin-flip interaction between the proton and neutron orbitals. The change in the energy gap between the {\it pf} orbitals leads to rapid changes in collectivity for nuclides along the N = 40 isotones. 
Intruder states in such cases can even become ground states due to strong quadrupole deformation, hence implying an IOI around N = 40 for Fe and Cr isotopes \cite{Santamaria2015, Porter2021}, similar to the N = 20 IOI in Mg \cite{Poves1987}. 

A recent precision mass measurement \cite{Mougeot2018} shows a gradual enhancement of collective behavior in Cr isotopes indicated by the flattening of the two-neutron separation energies up to N=39. However, calculations with the Universal nuclear energy-density functional (UNEDF0) interaction \cite{Kortelainen2010} performed for comparison predict a spherical $^{64}$Cr (N=40) ground state with an onset of deformation beginning only at $^{68}$Cr ($N=44$). Santamaria {\em et al.}\cite{Santamaria2015} presents a similar flat behavior of the E$_{2+}$ energies from N=38 to N=42 for Cr isotopes. Their work suggests the maximum collectivity has not been reached at $^{66}$Cr and shows the need for further measurements to elucidate the extension from N=40 IOI to $N=50$ isotones. A new calculated level scheme for $^{62, 64}$Cr beyond the (4$_1^+$) state with $\gamma$-ray spectroscopy was recently proposed \cite{Gade2021}. Their work predicts collective structures for these isotopes and argues $^{62}$Cr to be a transitional system to the path to maximal collectivity of the N=40 IOI. 

Here, we present results from precise measurements of $^{59, 61-65}$Cr with a Multiple-Reflection Time-of-Flight Mass Spectrometer (MR-TOF-MS), including the first high-precision direct mass measurements of $^{64, 65}$Cr. Our mass measurement of $^{64, 65}$Cr extends the mass trend of chromium isotopes and sheds light on how the collectivity evolves in this mass region with their comparison to new results from state-of-the-art ab-initio valence-space in-medium similarity renormalization group (VS-IMSRG) calculations \cite{Miyagi2020}, with which we explore the entire N=40 island of inversion region from calcium to nickel.

\section{Experiment}

\begin{figure}[tb]
\centering
  \includegraphics[width=\linewidth]{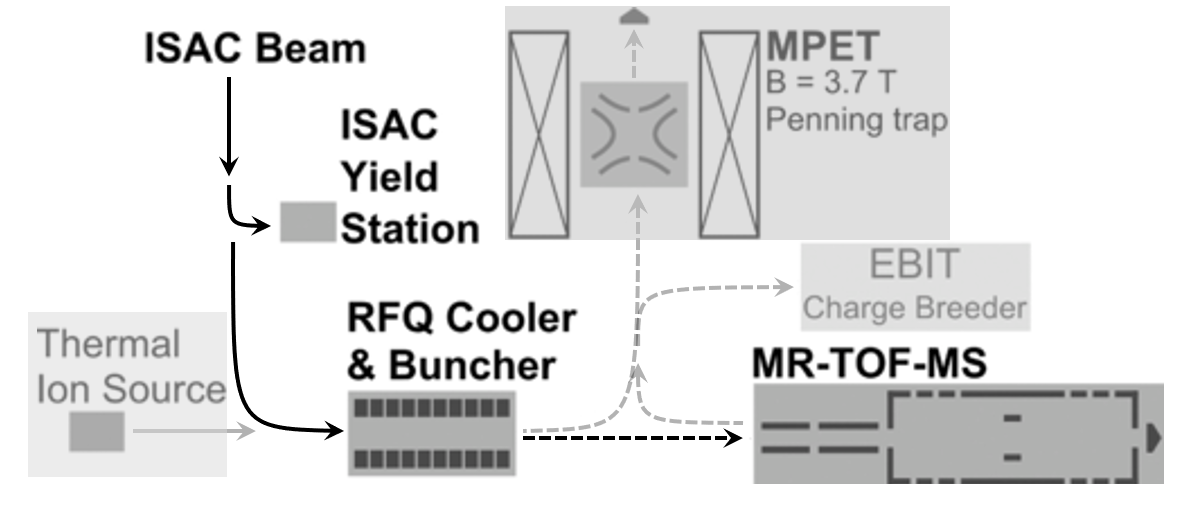}
\caption{A schematic diagram of TITAN showing various ion traps. The MR-TOF-MS was used during the measurements and the traps that were not used are shaded in gray.}
\label{fig:fig1}
\end{figure}

The mass measurements for $^{59, 61-65}$Cr were performed with the MR-TOF-MS \cite{Jesch2015}, part of the TITAN facility shown in Fig. \ref{fig:fig1}. A 480 MeV, 15 $\mu$A proton beam was impinged upon a UC$_X$ target at the Isotope Separator and Accelerator (ISAC) facility \cite{Ball2016}. The element of interest $_{24}$Cr along with few contaminant atomic species and diatomic oxide and fluoride molecules were surface ionized by a heated Re surface ion source. The ionization of $_{24}$Cr was enhanced using a two-step resonant ionization laser scheme by the TRIUMF Resonant Ionization Laser Ion Source (TRILIS) \cite{Lassen2017, Kudryavtsev2014}. 

The resulting beam was then reduced to a single mass unit using the ISAC dipole magnetic separator \cite{Bricault2002} and was transported to the Radio-Frequency Quadrupole (RFQ) cooler $\&$ buncher \cite{Brunner2012} at the TITAN facility. Here, it was bunched and cooled for 20 ms by collisions with He buffer gas. The cooled radioactive beam was then sent to the TITAN MR-TOF-MS \cite{Jesch2015} for mass measurements. 

Details of the operation and performance of the TITAN MR-TOF-MS are summarised in \cite{Reiter2021}. The device is based on the MR-TOF-MS in operation at the FRS Ion Catcher \cite{Wolfgang2013}. Inside the MR-TOF-MS, the ion bunches were transported to a dedicated injection trap, from which they were injected into a time-of-flight mass analyzer, consisting of nine electrodes (four electrostatic mirror electrodes on each side of a grounded drift tube electrode) \cite{YAVOR20151}. The ion bunches underwent around $600$ turns between the electrostatic mirrors to increase the flight path, which increased the mass resolution to nearly $300\,000$ FWHM. After a predetermined number of turns, the potential of the second mirror was lowered and the ions were impinged on a MagneTOF detector to record the time-of-flight spectrum. A mass range selector (MRS) \cite{Dickel2015} in the MR-TOF-MS analyzer, comprised of two pairs of deflecting electrodes at the center of the central drift tube, was employed to ensure an unambiguous mass spectrum. 

\begin{figure}[tb]
\centering
\begin{minipage}[t] {\linewidth}
\includegraphics[width=\linewidth]{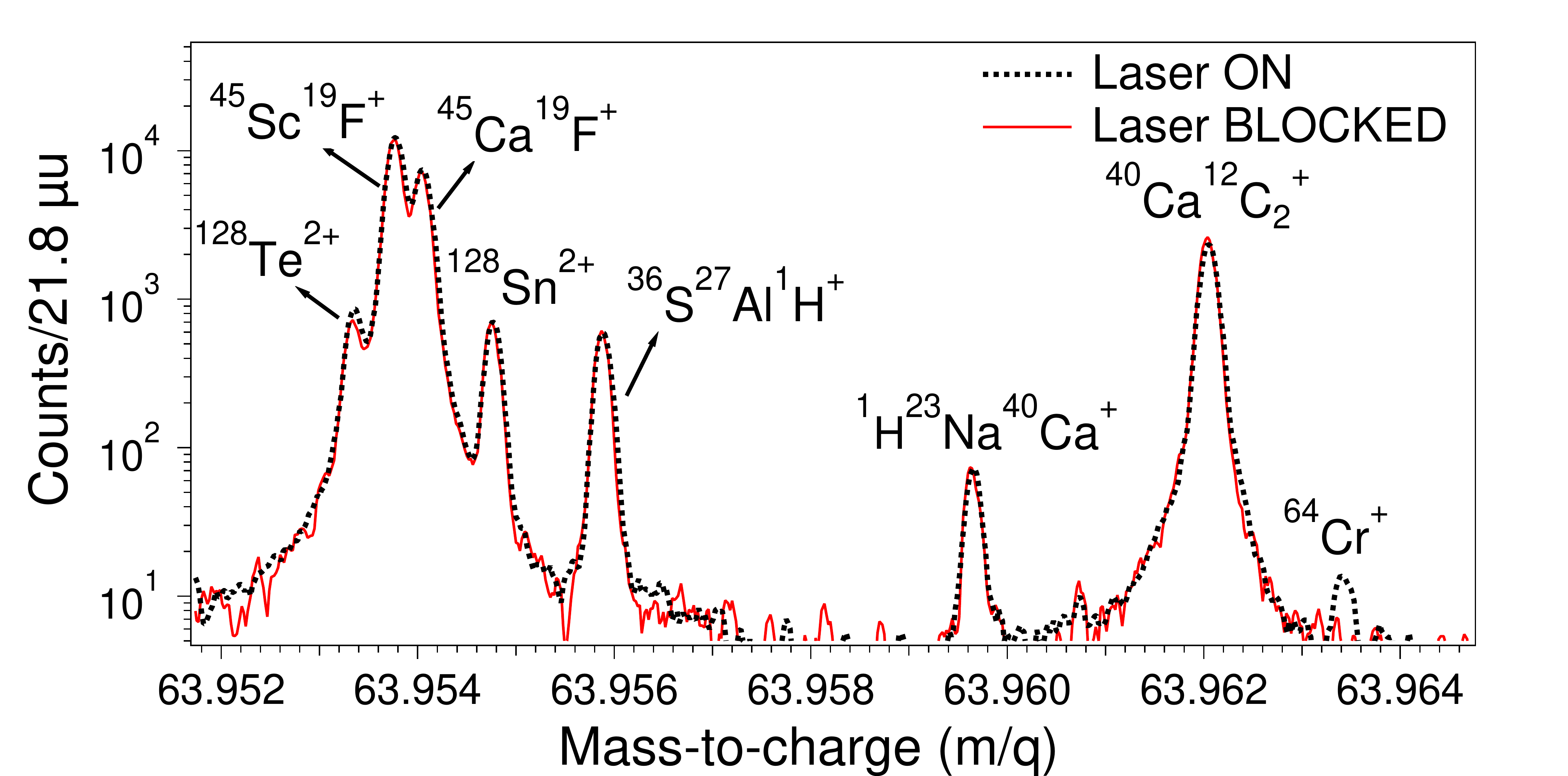}
\end{minipage}

\hspace*{\fill} 

\begin{minipage}[t] {\linewidth}
\includegraphics[width=\linewidth]{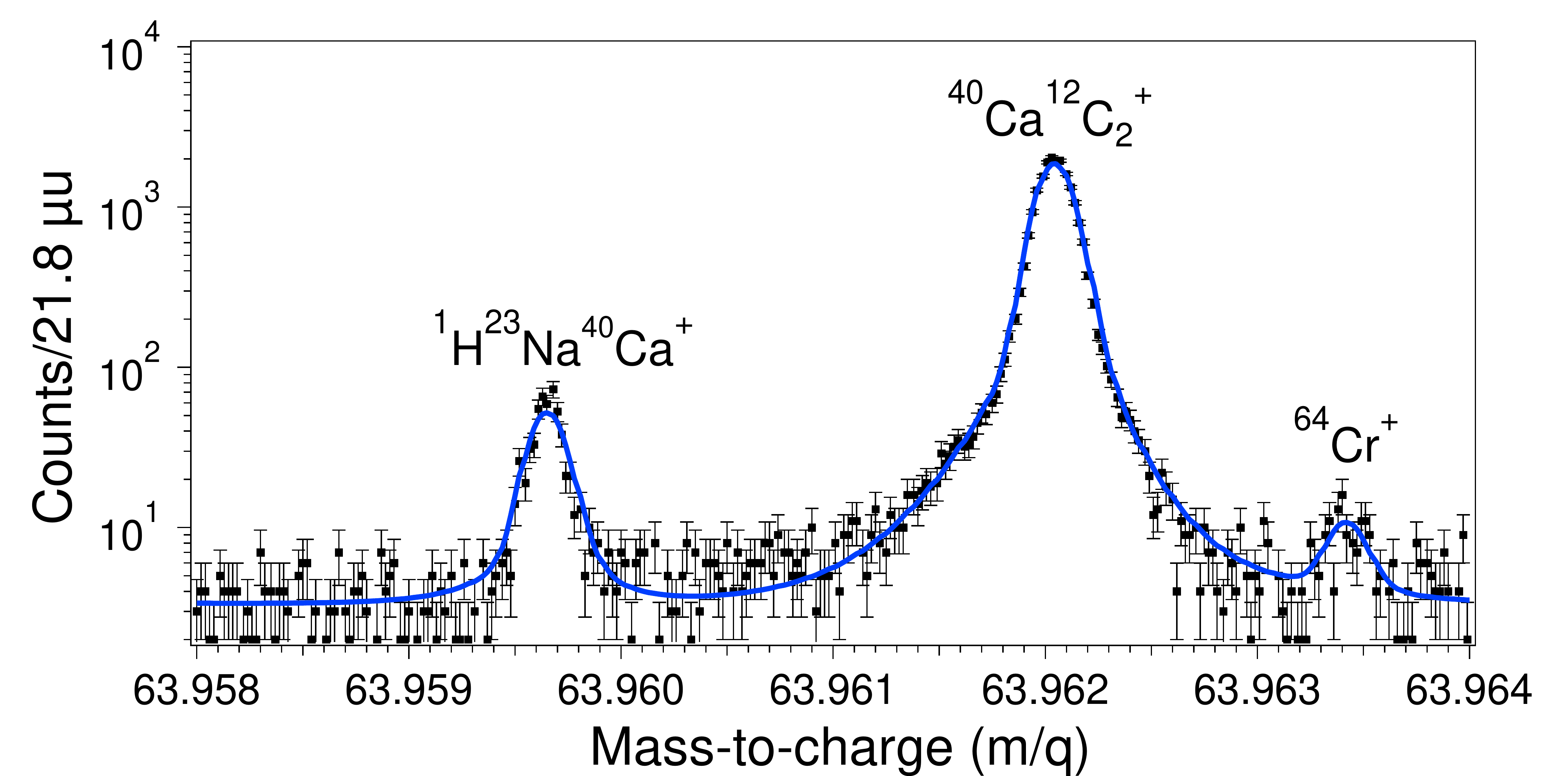}
\end{minipage}
\caption{(Top) Laser on (black dashed) versus blocked (red solid) spectra for $^{64}$Cr$^+$ shows a clear drop in ion counts when the laser was blocked and (Bottom) A zoomed spectra with error bars added shows the hyper-EMG fitting for $^{64}$Cr$^+$.}
\label{fig:fig2}
\end{figure}

The Cr yield at the MR-TOF-MS dropped with increasing neutron number from $\approx$ $10^3$ pps for $^{59, 61}$Cr to $<$ 1 pps for $^{64-65}$Cr, and is shown in Table \ref{table:table1}. Sufficient statistics for each mass allowed for the peak identification, where lower yield isotopes were collected for longer time. The beam contained $_{27}$Co$^{+}$, $_{28}$Ni$^{+}$ and $_{31}$Ga$^{+}$ and doubly charged $_{50}$Sn$^{2+}$, $_{51}$Sb$^{2+}$, $_{54}$Xe$^{2+}$ and $_{56}$Ba$^{2+}$ atoms. Dominant diatomic oxides and fluorides present were TiO$^+$, CaF$^+$, and ScO$^+$ as seen in Fig. \ref{fig:fig2}. Verification of the $^{59,  61-65}$Cr mass peaks was performed by blocking one step of the resonant ionisation laser. In this situation, the Cr intensity reduced by a factor of 2-7 for the $^{59,61-65}$Cr isotopes. Fig. \ref{fig:fig2} shows the drop in intensity of the $^{64}$Cr peak with blocked lasers, while the intensity of other peaks remained constant. For $^{65}$Cr, further investigation was required for the mass identification due to the lower statistics and a smaller change in the signal intensity with and without the resonant laser ionization. A reduction in laser enhancement has commonly been observed at ISAC for the most short-lived isotopes.

\begin{figure}[tb]
\centering
\includegraphics[width=\linewidth]{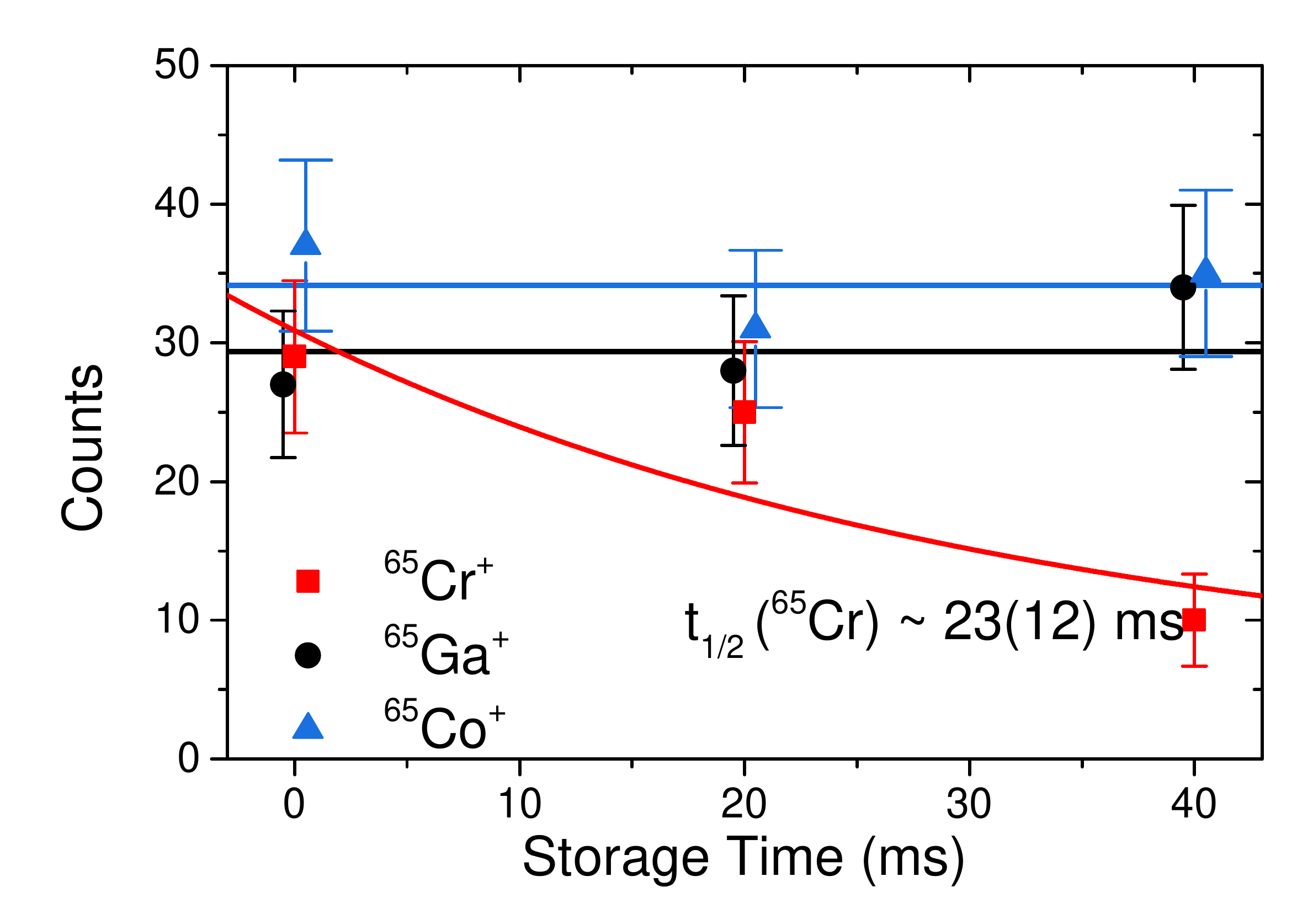}
\caption{Number of ions as a function of storage time in the MRTOF RFQ trap for $^{65}$Cr$^+$, $^{65}$Ga$^+$, and $^{65}$Co$^+$. A half-life of 23 $\pm$ 12 ms was obtained for $^{65}$Cr compared to the reported NUBASE value of 27.5 $\pm$ 2.1 ms.
The $^{65}$Cr$^+$ counts drops nearly exponentially while the $^{65}$Ga$^+$ and $^{65}$Co$^+$ counts are constant.}
\label{fig:fig3}
\end{figure}

To validate the presence of $^{65}$Cr, we performed a storage analysis measurement where the ion bunches were stored for varying times in the internal RFQ beam line of the MR-TOF-MS before transporting them to the mass analyzer \cite{Mukul2020}. These storage times were chosen with respect to $^{65}$Cr's half-life of 28 ms. From Fig. \ref{fig:fig3}, it can be seen that the intensities of $^{65}$Cr dropped while the intensities of long-lived $^{65}$Co and $^{65}$Ga stayed constant. A half-life of 23(12) ms could be determined for $^{65}$Cr which is consistent with the literature value of 27.5 (2.1) ms, further validating our assignment of $^{65}$Cr.

The recorded TOF spectra were converted to the mass spectra using a calibration function
\begin{equation}
m/q = c(t-t_0)^2.
\label{eq:eq1}
\end{equation}

Here, {\it m/q} is the mass-to-charge ratio, {\it c} and {\it $t_0$} are the calibration parameters, and {\it t} is the time of flight between injection into the analyzer and detection. The parameter {\it $t_0$} is a small offset in timing caused by electronic delays and was determined from the time of flight of {$^{39, 41}$K} peaks undergoing only a time-focus shift turn \cite{Reiter2021}. It amounted to $209.2(5)$~ns. Parameter {\it c} was derived for each mass peak from a well-known reference ion. A time-resolved calibration (TRC) \cite{Ebert2016} was performed to correct for any time-dependent drifts arising from power supply instabilities and temperature fluctuations using the mass data acquisition (MAc) software package \cite{Dickel2017}. The dominant CaF$^+$ and TiO$^+$ ions were chosen as the calibration species for the TRC of $^{59}$Cr$^+$, $^{61}$Cr$^+$ and $^{62}$Cr$^+$ -- $^{64}$Cr$^+$ masses, respectively. For $^{65}$Cr$^+$, $^{65}$Cu$^+$ was used as a reference for TRC. To determine the peak centroids, a hyper-EMG function \cite{Purushothaman2017} was fitted using the \texttt{emgfit} Python wrapper \cite{emgfit}. The hyper-EMG function particularly excels in fitting overlapping peaks such as the case of $^{64}$Cr$^{+}$ as can be seen in Fig. \ref{fig:fig2} \cite{Ayet2019}. The uncertainty in the determined mass values is a combination of the fitting uncertainty, AME uncertainty of the reference/calibrant ion, and a systematic uncertainty $\delta m/m$ of 1 $\times$ 10$^{-7}$ \cite{Jacobs2019, Paul2021}. On average, there is less than one ion per cycle in the MR-TOF. Therefore, the systematic shifts due to ion-ion interaction during this beam time were negligible. The weighted mean of the deviation of all identified species in the spectra, both atomic and molecular, from AME2020 masses was at the keV level, validating our assigned uncertainty.

\section{Results and Discussions}

\begin{table*}[t]
\centering
\caption{ Mass excess values (in keV) determined in this work in comparison to previous works is shown for $^{59,61-65}$Cr masses. Values from the ISOLTRAP \cite{Mougeot2018} and the NSCL TOF-B$\rho$ measurements \cite{Meisel2016, Meisel2020} were taken for comparison with our results as well as for the reference/calibrant ions. We also list the approximate yield at the ISAC yield station and the total counts in MR-TOF for this measurement.}
\label{table:table1}
\begin{tabular}{ccccllcccc}
Isotopes & Yield(pps)  & Counts & Mass Calibrant & \multicolumn{3}{c}{Mass   Excess (keV)}                                                                                                          \\  \cline{5-7}
                  &              &           &                               & \multicolumn{1}{c}{This work}      & \multicolumn{1}{c}{Ref \cite{Mougeot2018}}   & \multicolumn{1}{c}{Ref. \cite{Meisel2016, Meisel2020}}                                 \\ \hline
$^{59}$Cr         &  17000       &     39000         & $^{59}$Co                       & -48113(7)  & -48115.9(0.7)  & -48540(440)               \\
$^{61}$Cr         &   -      &     300         & $^{45}$Sc$^{16}$O                       & -42497(14) & -42496.5(1.8)    & -43080(510)                 \\
$^{62}$Cr         &     850       & 170                 & $^{30}$Si$^{32}$S                       & -40855(19) & -40852.6(3.5)       &   -40890(490)       \\
$^{63}$Cr         & 50        &190           & $^{44}$Ca$^{19}$F                       & -36204(18) & -36178(73)    & -35940(430)                   \\
$^{64}$Cr         & 1     &170               & $^{48}$Sc$^{16}$O                       & -33573(26) &       &   -33640(300)            \\
$^{65}$Cr         & 0.3    &140            & $^{46}$Ca$^{19}$F                       & -28208(45) &   & -27280(780)       
\end{tabular}
\end{table*}

Table~\ref{table:table1} lists the mass excess values evaluated in this work as well as comparisons with the previous measurements \cite{Mougeot2018, Meisel2016, Meisel2020}. The recent AME2020\cite{AME2020} evaluation include the measurements from \cite{Mougeot2018}. The TOF-B$\rho$ measurements for $^{65}$Cr reported in \cite{Meisel2016, Meisel2020} was replaced in the AME2020 evaluation by an extrapolated value judged to be more accurate. This highlights the need for not only precise, but also accurate measurements. Here, mass excess {\it (ME)} is defined as, {\it ME(N, Z) = m$_a$(N, Z) - Am$_u$}, where {\it m$_a$} is the measured atomic mass and {\it m$_u$} = 1 u = 9.314940954(57) $\times$ 10$^8$ eV/$c^2$. The reference ions strongest in statistics were chosen as calibrants for the mass determination and are listed for each chromium mass in Tab.~\ref{table:table1}.

\begin{figure}[tb]
\centering
\includegraphics[width=\linewidth]{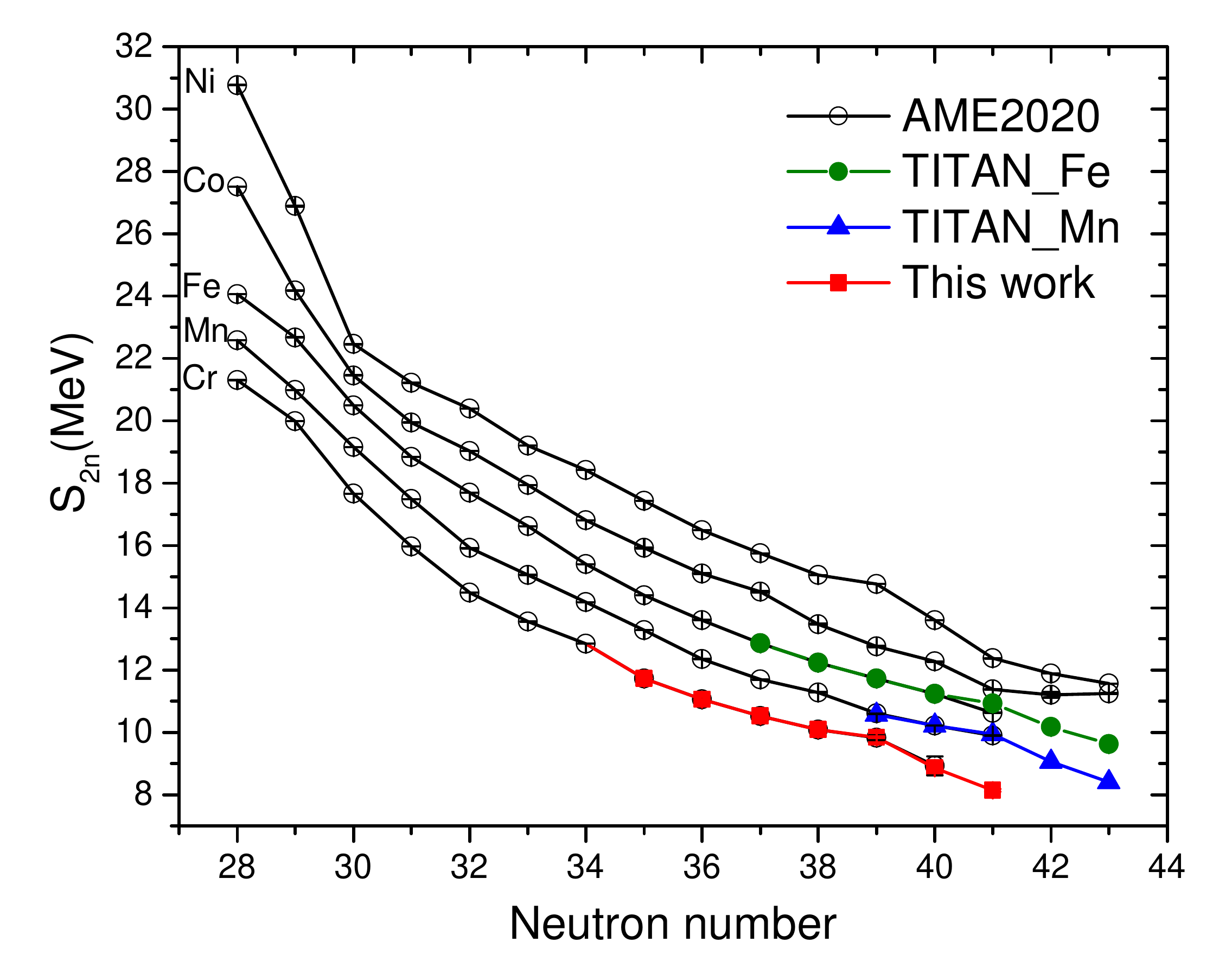}
\caption{Two-neutron separation energy (in MeV) for Ni (Z=28) through Cr (Z=24) is shown including the comparisons with AME2020 \cite{AME2020}. For Fe and Mn, recent TITAN measurements are also shown \cite{Gallant2021, Porter2021}.}
\label{fig:fig4}
\end{figure}

\begin{figure}[tb]
\centering
\includegraphics[width=\linewidth]{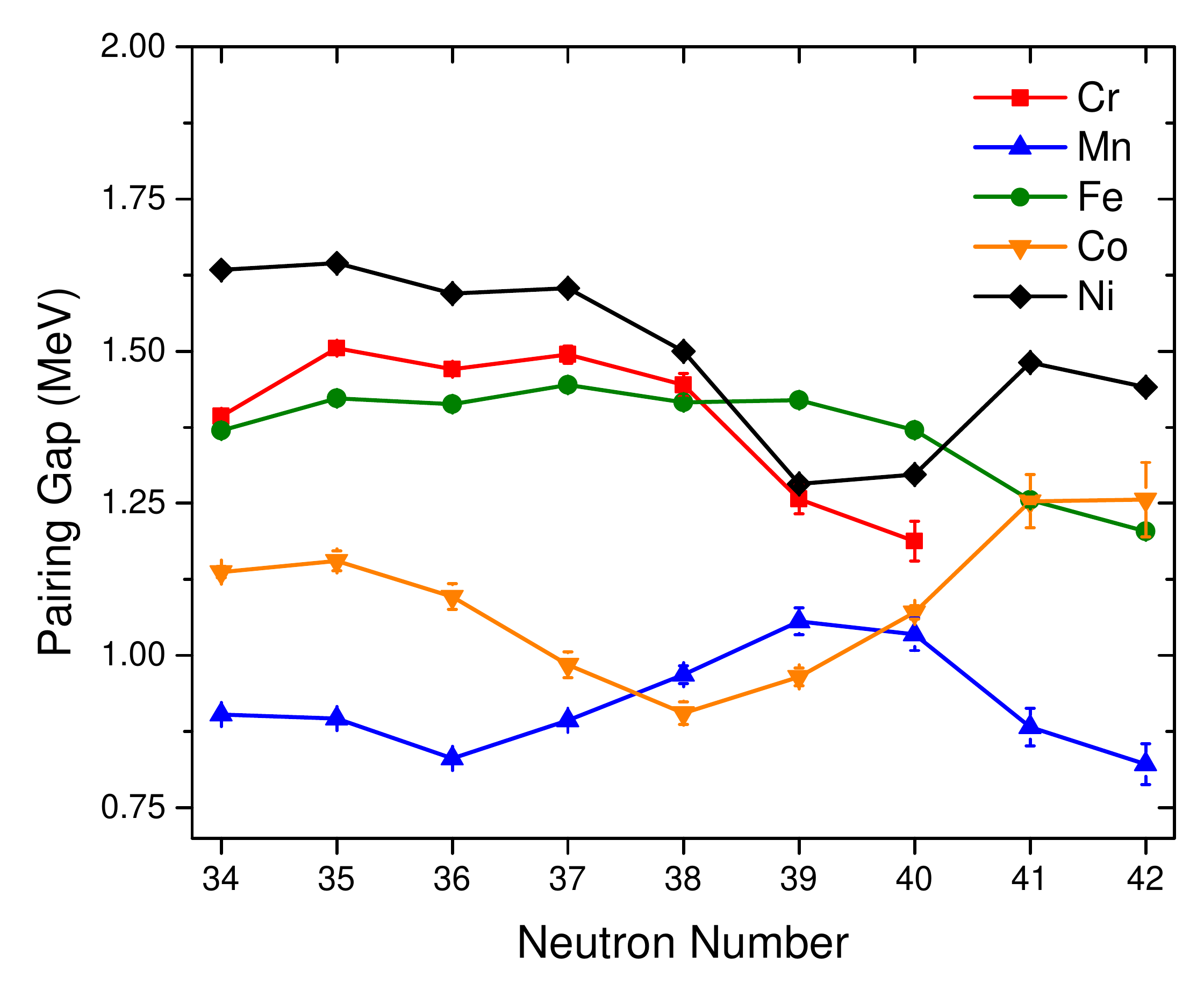}
\caption{{Pairing gap energies along the N=40 IOI for Cr (Z=24) through Ni (Z=28) chain.}}     
\label{fig:fig6}
\end{figure}

Based on the new mass values we calculate the two-neutron separation energies {\it S$_{2n}$}, shown in Fig.~\ref{fig:fig4}, 
defined by Eqn.~\ref{eq:eq2} for Cr (Z=24) to Ni (Z = 28). Here $ME(N,Z)$ is the mass excess of an isotope with N neutrons and Z protons, $M_n$ is the neutron mass excess. 
\begin{equation}
S_{2n}(N,Z) = ME(N-2,Z) - ME(N,Z) + 2M_n
\label{eq:eq2}
\end{equation}
The {\it S$_{2n}$} trend near the well-known shell-closure at N=28 is shown for comparison to the signatures at the N=40 region.

Our results agree with and extend beyond \newline Mougeot {\em et al.}'s precise measurements of $^{59-63}$Cr \cite{Mougeot2018}. Their {\it S$_{2n}$} trend shows that unlike a sudden onset of deformation towards N=40 \cite{Meisel2016, Naimi2012, Babcock2016}, an upward curvature in {\it S$_{2n}$} exists for the Cr chain up to N=39, which is interpreted as a gradual onset of collectivity in the ground-state of neutron-rich Cr isotopes similar to the neighboring Mn and Fe chains. Our extended measurements for $^{64,65}$Cr isotopes show that the {\it S$_{2n}$} values do not follow this upward trend when approaching N=40. This also contradicts the similarity in the {\it S$_{2n}$} of Cr and Mn chain past $^{63}$Cr (N=39).

In order to better understand the nature of the deformation in
the Cr isotopes across N=40, Fig.~\ref{fig:fig6} shows the pairing gap energies for Cr (Z=24) through Ni (Z=28), defined as
\begin{dmath}
P_{n}(N,Z) = \frac{(-1)^{N+1}}{4}(S_n(Z, N+1) \\
- 2S_n(Z,N) + S_n(Z,N-1)).
\end{dmath}

The pairing-gap energies in Cr shows a nearly constant behavior until N=37, but then drops rapidly approaching N=39, which further points towards a more rapid change in correlation effects in Cr towards N=40. Interestingly, the Cr chain seems to tail the behavior seen in the Ni chain, whereas the in-between Mn chain shows a maximum paring-gap energy at N=39. In Fe the constant paring-gap energy region expands beyond N=39, but drops to the energy observed near the minimum in Ni only at N=42, and in Co a reduction already occurs already past N=35. Similar discussions have been made by \cite{Santamaria2015} and have indicated the delicate interplay between quadrupole deformation and pairing correlations.

\begin{figure}[tb]
\centering
\includegraphics[width=\linewidth]{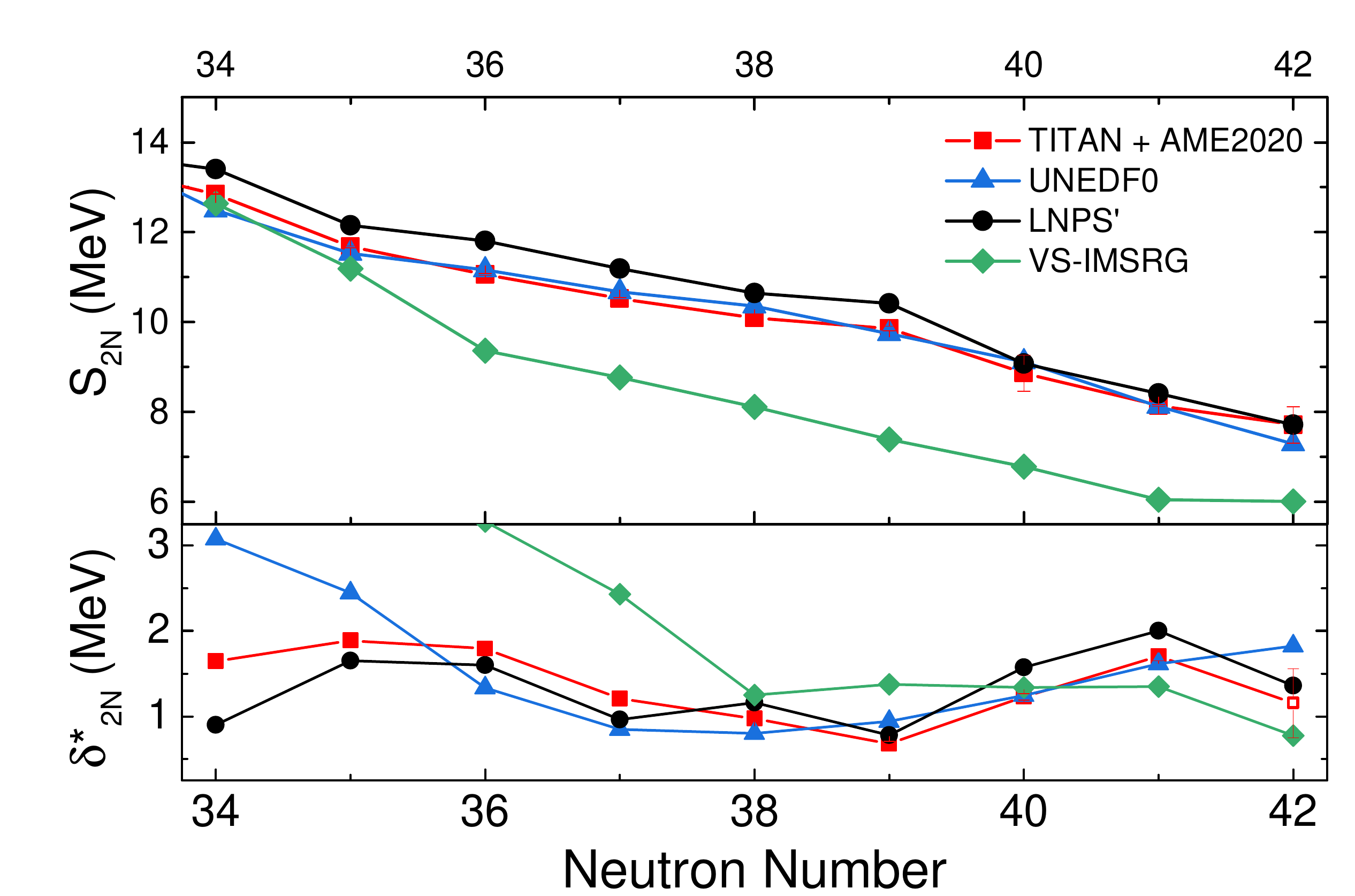}
\caption{Comparison of S$_{2n}$ values (top) and $\delta$S$_{2n}$ values (bottom) from VS-IMSRG($pf_{5/2}g_{9/2}d_{5/2}$), LNPS' and UNEDF0 theories with the TITAN results is shown.}
\label{fig:fig5}
\end{figure}

To shine more light on the behavior observed in the Cr chain, we performed theoretical calculations using different approaches, as shown in Fig.~\ref{fig:fig5}.    
In addition to {\it S$_{2n}$} we shows the two-neutron shell gap parameter, {\it $\delta^*_{2n}$} in MeV
defined by 
\begin{equation}
\delta^*_{2n}(N,Z) = S_{2n}(N-2,Z) - S_{2n}(N,Z).  
\label{eq:eq3}
\end{equation}
As seen in the figure, the decrease in the magnitude of {\it S$_{2n}$} agrees with the results from the mean-field calculations of the odd-even and even-even Cr isotopes from the UNEDF0 theory \cite{Kortelainen2010}. However, the $\delta^*_{2n}$ trend does not agree with the UNEDF0 theory. Both our {\it S$_{2n}$} and $\delta^*_{2n}$ trends match quite well with the LNPS' calculations, a modified version of LNPS \cite{Meisel2016} with 30 keV more attractive global monopole term \cite{Mougeot2018}. Previous results from LNPS calculations \cite{Mougeot2018, Meisel2016, Lenzi2010} that include a valence space based on a $^{48}$Ca core, a full {\it pf}-shell for protons and the neutron {\it pf}$_{5/2}$g$_{9/2}$d$_{5/2}$ orbitals, have shown a significant improvement over the GXPF1A phenomenological calculations, and highlights the need to include the g$_{9/2}$ and d$_{5/2}$ neutron states to properly address the shell structure.

Fig.~\ref{fig:fig5} also shows comparisons with a recently developed ab initio approach based on the VS-IMSRG, but which extends the calculations of the chromium chain with a mixed-parity valence space~\cite{Miyagi2020}. In this theory, an $A$-body unitary transformation is constructed to decouple the excitations involving the outside of a valence space~\cite{Herg16PR,Morr15Magnus,Bogner2014,Stroberg2019}. From the computational point of view, the actual unitary transformation is constructed with the truncation to two-body level and induced many-body terms are neglected [referred to as the VS-IMSRG(2) approximation]. Following the framework outlined in Ref.~\cite{Stroberg2017}, the energies of Cr isotopes are calculated starting from the 1.8/2.0 (EM) interaction~\cite{Hebeler2011}, which provides the accurate ground-state energies compared to experiment up to at least the $^{132}$Sn region~\cite{Stroberg2021,Miya21Heavy}, and in particular separation energies and shell gaps in this region~\cite{Leis18Ti,Mougeot2018,Xu19Sc,Leis21Sc,Porter2021}. 

In this work, starting from the $\hbar\omega=16$ MeV harmonic oscillator basis, our calculations are performed in the 15 major shell space to derive the effective Hamiltonian in the proton $pf$ and neutron $pf_{5/2}g_{9/2}d_{5/2}$ valence space above $^{48}$Ca core.
Note that the neutron $d_{5/2}$ orbital is added to the valence space here, in contrast to the calculations in Ref.~\cite{Mougeot2018}. The three-body interaction matrix elements are restricted up to $E_{\rm 3max}=16$, where $E_{\rm 3max}$ is defined as the sum of the three-body oscillator quanta. The center-of-mass Hamiltonian coefficient $\beta=3$ is used in the following discussion and we observed that the $\beta$-dependence is negligibly small.
For N$>$35, the calculated $S_{2n}$ are lower than the experiment, which was already observed in Ref.~\cite{Mougeot2018}. The discrepancy is attributed to the many-body correlations beyond the VS-IMSRG(2) approximation, implying the enhancement of the collectivity in N$>$35. 

It should be noted that the VS-IMSRG results systematically overestimate the excitation energies, while the trend of the experiment is well reproduced as shown in earlier works~\cite{Miyagi2020, Simonis2017}. While too strong for N$ < $38, the VS-IMSRG {\it $\delta^*_{2n}$} results follow the experimental trends in the IOI region well, as shown in Fig.~\ref{fig:fig5}, and thus we expect that the underlying physics is mostly captured through the VS-IMSRG(2) approximation. Nevertheless, since collectivity is an inherently many-body effect, we expect further improvement when advancing to the IMSRG(3) truncation for such calculations. Indeed implementation of IMSRG(3) has already been carried out in the single-reference formulation of the IMSRG for closed-shell systems~\cite{Heinz21IMSRG3}, and first results for the VS-IMSRG are currently in preparation.

\begin{figure}[tb]
\centering
\includegraphics[width=\linewidth]{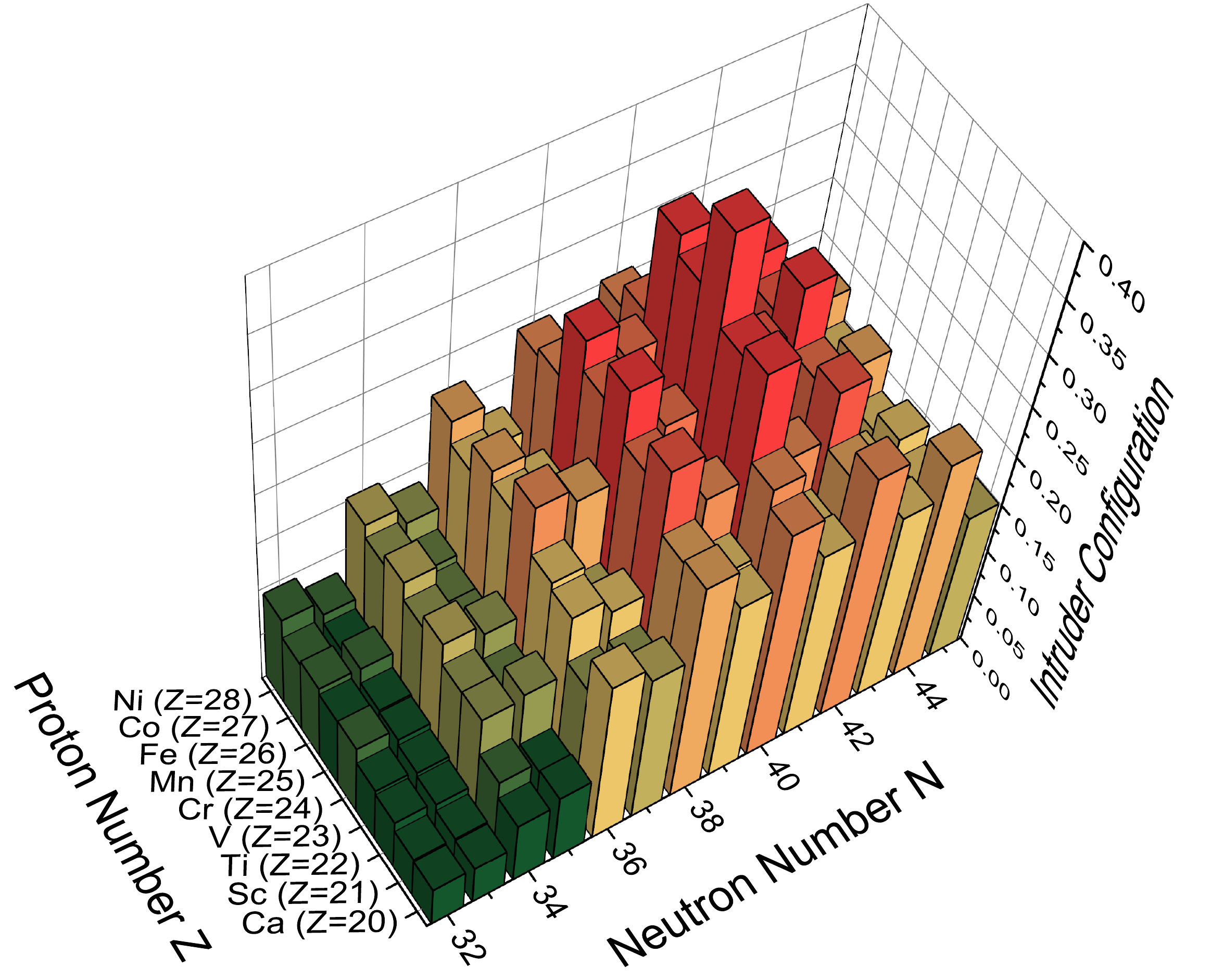}
\caption{The intruder contribution of the calculated ground-state wave function for the elements Ca (Z=20) through Ni(Z=28), computed by the VS-IMSRG(2) within the proton $pf$ neutron $pf_{5/2}g_{9/2}$ valence space.} 
\label{fig:fig7}
\end{figure}

To further examine the extend of the N=40 IOI in the neighboring elements, we expand the VS-IMSRG calculations to cover the full region between Ca and Ni. In Fig.~\ref{fig:fig7}, the contributions of intruder configuration for Cr and neighboring elements are visualized using the calculated ground-state wave function from the VS-IMSRG within the proton $pf$ and neutron $pf_{5/2}g_{9/2}$ valence space  above $^{48}$Ca.
As seen in Fig. ~\ref{fig:fig7}, towards N=40, the intruder configuration increases in the Cr ground states, a sign of IOI behavior at N=40~\cite{Lenzi2010}. Also, the number of neutrons across the gap is maximized at N=40, consistent with the LNPS calculations~\cite{Lenzi2010}.

This is similar to the trend in the {\it f$_{7/2}$} and {\it p$_{3/2}$} orbitals for $^{32}$Mg~\cite{Lungjvall2010, Miyagi2020}. Additionally, it is seen that the intruder state configuration is strongest for Cr compared to the neighboring nuclei, such as Mn, Fe and Co, shown in Fig. \ref{fig:fig7}, implying that $^{64}$Cr is the pinnacle of the N=40 IOI. The next strong contribution is seen for Mn and Fe, with a similar trend where the intruder contribution is the strongest at N=40 followed by N=42. For Fe, this result is in agreement with the maximum observed quadrupole deformation between $^{66}$Fe and $^{68}$Fe (i.e. N = 40 and N = 42) from the mean-field calculations \cite{Porter2021}. Another recent work on Mn have shown a high pairing gap approaching N=40 attributed to neutron occupying higher lying orbitals \cite{Gallant2021}. These results are in line with the results from LNPS calculations \cite{Lenzi2010} that show the onset of deformation at N=40 in the Fe chain and at N=38 in Cr isotopes. 

\section{Conclusion and Discussion}
Our results reduce the uncertainties of the mass measurements for $^{63-65}$Cr masses. Moreover, this presents the first high-precision direct mass measurements of $^{64, 65}$Cr. Our results provide compelling evidence for maximal collectivity at $^{64}$Cr centering it in an island of inversion around N=40. Comparison of our results with global mass-models and the ab-initio VS-IMSRG calculations indicates the dominance of the intruder configurations for $^{64}$Cr, and shows the overall trend of collectivity in the region. The precise data provides important constraints to guide the ongoing development of ab-initio approaches to nuclear structure. High precision mass measurements beyond N=41 would be required to more fully probe the N=40 island of inversion. Advancement in the ab initio theory in this region will be significant in illuminating the nuclear shell structure for the $Z=24$ nuclei. 

\section{Acknowledgement}

The authors would like to thank M. Good for his continual support, J. Lassen and the TRILIS group for their help with setting up the Cr lasers scheme for signal optimization, and  S. R. Stroberg for the imsrg++ code~\cite{Stro17imsrg++} used to perform these calculations. The shell-model diagonalization of VS-IMSRG calculations were done with the KSHELL code~\cite{Shimizu2019}.
This work was supported by the Natural Sciences and Engineering Research Council (NSERC) of Canada under grants SAPIN-2018-00027 and RGPAS-2018-522453, by the National Research Council (NRC) of Canada through TRIUMF, by the UK Science and Technology Facilities Council STFC Grant No: ST/V001051/1, and by the Deutsche Forschungsgemeinschaft (DFG, German Research Foundation) -- Project-ID 279384907 -- SFB 1245. Computations were performed with an allocation of computing resources on Cedar at WestGrid and Compute Canada, and on the Oak Cluster at TRIUMF managed by the University of British Columbia department of Advanced Research Computing (ARC). We also thank K. Sieja for providing the LNPS' results.

\bibliography{mybibfile}

\end{document}